\documentstyle[aps,preprint]{revtex}
\input{epsf}
\tighten

\begin{document}

\draft

\title{Simple relations among $E2$ matrix elements of low-lying
collective states} 

\author{V. Werner$\,^1$, P. von Brentano$\,^1$ and R.V. Jolos$\,^{1,2}$}

\address{$^1\,$ Institut f\"ur Kernphysik, Universit\"at zu K\"oln, 
                50937 K\"oln, Germany}
\address{$^2\,$ Bogoliubov Laboratory, Joint Institute for Nuclear Research, 
                141980 Dubna, Russia}

\date{\today}

\maketitle

\begin{abstract}
A method is developed to derive simple relations among the
reduced matrix elements of the quadrupole operator between
low-lying collective states. As an example, the fourth order scalars 
of $Q$ are considered. The accuracy and validity of the proposed
relations is checked for the ECQF Hamiltonian of the IBM--1 in the
whole parameter space of the Casten triangle. Furthermore these
relations are successfully tested for low-lying collective states in
nuclei for which all relevant data is available.
\end{abstract}


\narrowtext

Microscopic shell model wave functions of collective nuclear states
need a huge configurational space. However, experimental data
indicates that there are comparably simple relations between the wave
functions of different collective states, including the ground
state. The wave functions of some excited states can be described by
actions of one-body operators on the ground state wave function with
good accuracy. In even-even nuclei, where the ground state is a
$0^+$ state, the first $2^+$ state is given by the quadrupole operator
$Q$  acting on the ground state. The generalization of this concept
has been named the $Q$-phonon approach
\cite{Sie94,Ots94,Shi01,Pie94,Pie95,Pie98,Pal98}. In this approach one
describes the low-lying collective positive parity states of even-even
nuclei in the basis of multiple $Q$-phonon excitations of the ground
state, $|0_1^+\rangle$,
\begin{equation}
\label{eq:qexpand}
|L^+,n\rangle = N^{(L,n)} (\underbrace{Q...Q}_{n})^{(L)}
|0_1^+\rangle \ .
\end{equation}
In the framework of the IBM-1 \cite{AriIac75,IacAri87} it has been
shown over the whole parameter space of the ECQF Hamiltonian
\cite{WarCas82,Lip85} that each of the wave vectors of the yrast
states can be described by only one multiple $Q$-phonon configuration
with good accuracy \cite{Sie94,Pie94,Pie95}, which has recently been
confirmed by microscopic calculations in \cite{Shi01}.

The $Q$-phonon approximation implies the existence of selection rules
for the matrix elements of the quadrupole operator. Thus, one finds
that $E2$ transitions between $Q$-phonon configurations, that differ
by more than one $Q$-phonon, are weak compared to transitions between
those configurations that differ by only one $Q$-phonon. During the
last years much data on $\gamma$-soft nuclei has been collected,
especially in the A=130 mass region, which support these selection
rules, e.g. \cite{Gad00b,Wer01}.

The $Q$-phonon structure of the low-lying collective states allows one to
obtain quadrupole shape invariants \cite{Kum72,Cli86,Jol97,Wer00} from
rather few data. As an example we consider fourth order scalars
obtained by coupling the four quadrupole operators in different
ways. One obtains several different expressions for the fourth
order quadrupole shape invariants in terms of only a few $E2$ matrix
elements. These expressions can be used to derive approximate values
of various observables, e.g., for the quadrupole moment of the first
$2^+$ state or the lifetime of the first excited $0^+$ state, from
more easily accessible nuclear data. Such information is desirable for
nuclei where complete experimental information about low-lying states
is not -- or not yet -- available, as e.g. nuclei which are produced
using rare isotope beams.

There are three possibilities to couple the quadrupole operators to
obtain fourth order scalars,
\begin{eqnarray}
\label{eq:q40}
q_4^{(0)} & = & \langle 0_1^+| (Q\cdot Q) (Q\cdot Q) |0_1^+\rangle \ ,\\
\label{eq:q42}
q_4^{(2)} & = & \langle 0_1^+| \left[
[QQ]^{(2)}[QQ]^{(2)}\right]^{(0)} |0_1^+\rangle \ ,\\
\label{eq:q44}
q_4^{(4)} & = & \langle 0_1^+| \left[
[QQ]^{(4)}[QQ]^{(4)}\right]^{(0)} |0_1^+\rangle \ ,
\end{eqnarray}
The notation $[\ldots]^{(L)}$ abbreviates the tensor coupling of two
operators to angular momentum $L$. These three scalars are
proportional to each other according to Dobaczewski, Rohozi\'nski and
Srebrny in \cite{Dob87}, if the $Q$-operators commute. Then one
obtains the relations
\begin{equation}
\label{eq:cinf}
q_4^{(0)} = \frac{7\sqrt{5}}{2} q_4^{(2)} = \frac{35}{6} q_4^{(4)} \ .
\end{equation}
In the IBM-1 the $Q$-operators do not commute. The effect of the
noncommutativity of the components of the quadrupole operators scales
however with $1/N$ and is therefore neglected in first order. Below,
we will check the accuracy of Eq. (\ref{eq:cinf}) in the framework of
IBM-1.

In order to do this we decompose the scalars into sums over reduced
matrix elements,
\begin{eqnarray}
q_4^{(0)} & = & \ \sum_{i,j,k} \langle 0_1^+||Q||2_i^+\rangle \langle
2_i^+||Q||0_j^+\rangle \nonumber \\
\label{eq:q40sum}
& & \times \langle 0_j^+||Q||2_k^+\rangle \langle
2_k^+||Q||0_1^+\rangle \ , \\
q_4^{(2)} & = & \frac{1}{5\sqrt{5}} \sum_{i,j,k} \langle
0_1^+||Q||2_i^+\rangle \langle
2_i^+||Q||2_j^+\rangle \nonumber \\
\label{eq:q42sum}
& & \times \langle 2_j^+||Q||2_k^+\rangle \langle
2_k^+||Q||0_1^+\rangle \ , \\
q_4^{(4)} & = & \frac{1}{15} \sum_{i,j,k} \langle
0_1^+||Q||2_i^+\rangle  \langle 2_i^+||Q||4_j^+\rangle \nonumber \\
\label{eq:q44sum}
& & \times \langle 4_j^+||Q||2_k^+\rangle \langle
2_k^+||Q||0_1^+\rangle \ .
\end{eqnarray}
Using Eqs. (\ref{eq:q40sum})-(\ref{eq:q44sum}), the quantities
(\ref{eq:q40})-(\ref{eq:q44}) have been calculated 
gridwise -- using the code {\sc Phint} \cite{phint} -- for $N$=10
bosons over the whole IBM-1 symmetry space spanned by the
ECQF-Hamiltonian \cite{WarCas82,Lip85}
\begin{equation}
\label{eq:hecqf}
H_{ECQF} = a \ \left[(1-\zeta) \ n_d - \frac{\zeta}{4N} \ Q\cdot
Q\right] \ .
\end{equation}
The ECQF Hamiltonian interpolates between the symmetry limits of the IBM-1
using two structural parameters, $\zeta$ and $\chi$. Here, $n_d$ is
the $d$-boson number operator and $N$ is the total boson number. The
parameter $a$ has no structural meaning as it sets an absolute energy
scale, and $Q$ is the CQF quadrupole operator, both in the Hamiltonian 
and the $E2$ transition operator,
\begin{equation}
\label{eq:qcqf}
1/e_B \ T(E2)=Q=s^+\tilde{d}+d^+ s +\chi [d^+\tilde{d}]^{(2)} \ ,
\end{equation}
depending on the structural parameter $\chi$, with $-\sqrt{7}/2 \le
\chi \le 0$; $e_B$ is the effective boson charge. 
The result of this calculation is a near proportionality
of the $q_4^{(0)}$, $q_4^{(2)}$ and $q_4^{(4)}$, in accordance with
Eq. (\ref{eq:cinf}).

In view of the selection rules of the $Q$-phonon scheme, the sums
(\ref{eq:q40sum})-(\ref{eq:q44sum}) reduce drastically. In the first
approximation the set of $E2$ matrix elements necessary for the
calculation of $q_4^{(n)}$ ($n=0,2,4$) reduces to the following matrix
elements,
\begin{eqnarray}
\label{eq:214i}
\langle 2_1^+||Q||4_i^+\rangle & \longrightarrow & i=1 \ , \\
\label{eq:212i}
\langle 2_1^+||Q||2_i^+\rangle & \longrightarrow & i=1,2 \ , \\
\label{eq:210i}
\langle 2_1^+||Q||0_i^+\rangle & \longrightarrow & i=1,2,3 \ .
\end{eqnarray}
The first three $0^+$ states are taken into account, because
the $0_{2,3}^+$-eigenstates of the ECQF Hamiltonian are mixtures of
two- and three-$Q$-phonon $0^+$ configurations. Of course, if there
are low-lying non-collective $0^+$ states, the ECQF $0^+_{2,3}$
eigenstates may refer to higher lying physical states. We have
introduced the short notation $0^+_{QQ}$ by means of
\begin{equation}
\langle 0^+_{QQ}||Q||J\rangle^2 = \langle 0^+_{2}||Q||J\rangle^2 + 
\langle 0^+_{3}||Q||J\rangle^2 \ . \\
\end{equation}

\begin{figure}[h]
\epsfysize 4.5cm
\epsfbox{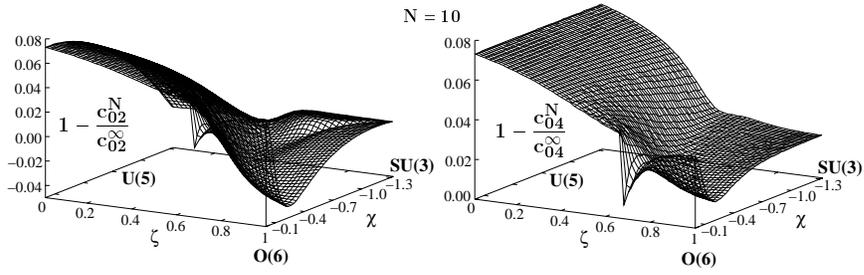}
\caption{The deviations of the factors $c_{02}^N$ and $c_{04}^N$ from the
limiting values, calculated gridwise over the whole ECQF symmetry
space for $N=10$ bosons.}
\label{fig:cij} 
\end{figure}

By using only the matrix elements (\ref{eq:214i})-(\ref{eq:210i}) in
(\ref{eq:q40sum})-(\ref{eq:q44sum}) we see that in each sum a
factor $\langle 0_1^+||Q||2_1^+\rangle^2$ appears, which may be
dropped, since we are interested in the
ratios. Eqs. (\ref{eq:q40sum})-(\ref{eq:q44sum}) now become:
\begin{eqnarray}
\label{eq:t40}
t_4^{(0)} & = & \ \langle 2_1^+||Q||0_1^+\rangle^2 + \langle
2_1^+||Q||0_{QQ}^+\rangle^2 \ , \\
\label{eq:t42}
t_4^{(2)} & = & \frac{1}{5\sqrt{5}} \left(\langle
2_1^+||Q||2_1^+\rangle^2 + \langle 2_1^+||Q||2_2^+\rangle^2\right) \ ,
\\
\label{eq:t44}
t_4^{(4)} & = & \frac{1}{15} \langle 2_1^+||Q||4_1^+\rangle^2 \ ,
\end{eqnarray}
where we use $t^{(n)}_4$ to distinguish these quantities from the
exact $q_4^{(n)}$ values.
These quantities are also approximately proportional to each other,
like the quantities $q_4^{(n)}$. For arbitrary values of the boson
number $N$ the $t_4^{(n)}$ values are related by factors $c_{0i}^N$
defined by
\begin{equation}
\label{eq:cdef}
t_4^{(0)} = \frac{1}{c_{02}^N} t_4^{(2)} = \frac{1}{c_{04}^N}
t_4^{(4)} \ ,
\end{equation}
which depend on the boson number and the dynamical symmetry character.
To obtain the values of the $c_{0i}^N$, we consider the $U(5)$, $SU(3)$
and $O(6)$ dynamical symmetry limits of the IBM-1 at first. 
The ECQF quadrupole operator (\ref{eq:qcqf}) is used, and one obtains
\begin{eqnarray}
\label{eq:t40limit}
t_4^{(0)} & = & \left\{         \begin{array}{c@{\qquad}c} 
                                7N-2 & U(5) \\
                                N(2N+3) & SU(3) \\
                                N(N+4) & O(6) 
                        \end{array} \right. \ , \\
\label{eq:t42limit}
t_4^{(2)} & = & \left\{         \begin{array}{c@{\qquad}c}
                                \frac{1}{\sqrt{5}} (2N+\chi^2-2) &
				U(5) \\
                                \frac{1}{28\sqrt{5}} (4N+3)^2 & SU(3)
				\\
                                \frac{2}{7\sqrt{5}} (N-1)(N+5) & O(6)
                        \end{array} \right. \ , \\
\label{eq:t44limit}
t_4^{(4)} & = & \left\{         \begin{array}{c@{\qquad}c}
                                \frac{6}{5} (N-1) & U(5) \\
                                \frac{6}{35} (N-1)(2N+5) & SU(3) \\
                                \frac{6}{35} (N-1)(N+5) & O(6) 
                        \end{array} \right. \ .
\end{eqnarray}
Comparing Eq. (\ref{eq:cdef}) and
Eqs. (\ref{eq:t40limit})-(\ref{eq:t44limit}) one obtains
proportionality factors for $N\rightarrow\infty$
\begin{equation}
\label{eq:climits}
c^{\infty}_{02} = \frac{2}{7\sqrt{5}} \quad , 
\quad c^{\infty}_{04} = \frac{6}{35} \ ,
\end{equation}
in agreement with the factors of
Eq. (\ref{eq:cinf}). These values hold also for finite $N$ in the $O(6)$
and the $SU(3)$ dynamical symmetry limits when one neglects $1/N^2$
terms. Only in the $U(5)$ limit a $1/N$ dependence is left, causing a
small deviation from the limiting values. The values of the parameters
$c_{0i}^N$ with finite $N$ differ slightly from those with
$N\rightarrow\infty$. Using Eqs. (\ref{eq:cdef})-(\ref{eq:t44limit})
we obtain improved relations in the dynamical symmetry limits
including the values of $c_{0i}^N$ for finite $N$. For nuclei far from
symmetries one can calculate the exact $c_{0i}^N$ using
Eq. (\ref{eq:cdef}) and interpolating in the IBM-1. We have done such
calculation for the IBM-1 using the ECQF-Hamiltonian
(\ref{eq:hecqf}). Fig. \ref{fig:cij} shows the values of the parameter
\begin{equation}
\label{eq:cdev}
1-\frac{c_{0i}^N}{c_{0i}^{\infty}} \quad , \quad (0i)=(02),(04)
\end{equation}
for $N$=10 bosons. Using the limiting values for $N\rightarrow\infty$
results in a systematical error below 10$\%$. We note that some
deviations arise from our use of only the $0^+_2$ state for the
$0^+_{QQ}$ configuration in this calculation.

From Eqs. (\ref{eq:t40})-(\ref{eq:cdef}) one obtains two relations for
the quadrupole moment of the $2^+_1$ state:
\begin{eqnarray}
\label{eq:q21amat}
\langle 2^+_1||Q||2^+_1\rangle^2 & + & \langle
2^+_1||Q||2^+_2\rangle^2 \nonumber \\
& = & 
\left(\frac{c_{02}^N}{c_{02}^{\infty}}\frac{c_{04}^{\infty}}{c_{04}^N}\right)
\cdot \frac{5}{9} \langle 2^+_1||Q||4^+_1\rangle^2 \ ,
\end{eqnarray}
\begin{eqnarray}
\label{eq:q21aeb}
Q_{2^+_1}^2 = \frac{32\pi}{35}\left[
\left(\frac{c_{02}^N}{c_{02}^{\infty}}\frac{c_{04}^{\infty}}{c_{04}^N}\right)\right.
& \cdot & B(E2;4^+_1 \rightarrow 2^+_1)
\nonumber \\
& - & \left. B(E2;2^+_2\rightarrow 2^+_1)\right]
\end{eqnarray}
and
\begin{eqnarray}
\label{eq:q21bmat}
& & \langle 2^+_1||Q||2^+_1\rangle^2 + \langle 2^+_1||Q||2^+_2\rangle^2
\nonumber \\
& & = \frac{10}{7}\cdot \frac{c_{02}^N}{c_{02}^{\infty}} \cdot
(\langle 2^+_1||Q||0^+_1\rangle^2 + \langle
2^+_1||Q||0^+_{QQ}\rangle^2) \ ,
\end{eqnarray}
\begin{eqnarray}
\label{eq:q21beb}
Q_{2^+_1}^2 & = & \frac{32\pi}{35} \left[ \frac{2}{7}\cdot
\frac{c_{02}^N}{c_{02}^{\infty}} \cdot [5
B(E2;2^+_1\rightarrow 0^+_1)\right. \nonumber \\
& + & \left. B(E2;0^+_{QQ} \rightarrow 2^+_1)] - B(E2;2^+_2
\rightarrow 2^+_1)\right] \ ,
\end{eqnarray}
where Eqs. (\ref{eq:q21amat}),(\ref{eq:q21aeb}) and
(\ref{eq:q21bmat}),(\ref{eq:q21beb}), respectively, differ only in
notation. In a first approximation with $c_{0i}^N/c_{0i}^{\infty}$=1,
and if we define $B(E2;2^+_1\rightarrow 2^+_1)\equiv 1/5 \langle
2^+_1||Q||2^+_1\rangle^2$, we can write expression (\ref{eq:q21amat})
in an intuitively interesting way:
\begin{equation}
B(E2;2^+_1\rightarrow 2^+_1) + B(E2;2^+_2\rightarrow 2^+_1) =
B(E2;4^+_1\rightarrow 2^+_1) \ .
\end{equation}
A relation similar to (\ref{eq:q21aeb}) for $N\rightarrow\infty$ has
been obtained in
\cite{Jol96}, but was derived in a much less transparent way and was
expressed using a rather difficult notation. Rewriting
Eq. (\ref{eq:q21beb}) we get a relation for $B(E2;0^+_{QQ} \rightarrow
2^+_1)$, and a second relation by inserting (\ref{eq:q21beb}) in
(\ref{eq:q21aeb}).

Extending our previous definitions \cite{Wer00} of quadrupole shape
invariants, we define now not only $K_4$, but $K_4^{(0)}$,
$K_4^{(2)}$ and $K_4^{(4)}$, depending on the coupling. We want to
obtain values, which characterize the nucleus and do not depend
strongly on the coupling scheme. Thus, with $q_2$=$\langle
0^+_1|Q\cdot Q|0^+_1\rangle$, we introduce
\begin{eqnarray}
\label{eq:k40}
K_4^{(0)} & = & \frac{q_4^{(0)}}{q_2^2} \ , \\
\label{eq:k42}
K_4^{(2)} & = & \frac{7\sqrt{5}}{2} \ \frac{q_4^{(2)}}{q_2^2} \ , \\
\label{eq:k44}
K_4^{(4)} & = & \frac{35}{6} \ \frac{q_4^{(4)}}{q_2^2} \ .
\end{eqnarray}
These quantities are all equal if the quadrupole operators commute. We 
note that in the large $N$ limit of the IBM-1 the theoretical values
for the $K_4^{(n)}$ are 1 for the $SU(3)$ and the $O(6)$, and 1.4 for
the $U(5)$ dynamical symmetry limit, distinguishing between
$\beta$-rigid and vibrational nuclei, respectively.
Applying the above results to the $K_4^{(n)}$
leads to an approximation formula for $K_4^{(0)}$ that has already been
obtained for $N\rightarrow\infty$ in \cite{Jol97},
\begin{equation}
\label{eq:k404}
K_4^{(0)}\approx \frac{7}{10}\frac{B(E2;4^+_1\rightarrow
2^+_1)}{B(E2;2^+_1\rightarrow 0^+_1)}\equiv K_4^{\rm appr.} \ .
\end{equation}
A second approximation is
\begin{equation}
\label{eq:k402}
K_4^{(0)}\approx \frac{7}{10} \left[
\frac{\frac{35}{32\pi}Q_{2_1^+}^2 +
B(E2;2^+_2\rightarrow 2^+_1)}{B(E2;2^+_1\rightarrow 0^+_1)}
\right] \ .
\end{equation}
Due to $K_4^{(0)}\in[1,1.4]$ it emerges from Eq. (\ref{eq:k402})
that, e.g., in the transition from $O(6)$ to $SU(3)$, where
$K_4^{(0)}$=$1$, the value of $Q_{2^+_1}^2/B(E2;2^+_1\rightarrow
0^+_1)$ rises from zero to 10/7, while the value of
$B(E2;2^+_2\rightarrow 2^+_1)/B(E2;2^+_1\rightarrow 0^+_1)$ drops from
10/7 to zero. Thus, these ratios characterize nicely the change of
structure.

In order to compare the relations with experimental data we
considered nuclei near dynamical symmetries, for which all needed data
is available. This data comes mostly from Coulomb excitation
experiments by D. Cline and co-workers \cite{Cli86,Cli99,Wu96,Sve95,Fa88}. In
Tables \ref{tab:tkerne},\ref{tab:qkerne} the results 
are given. The Os and Pt nuclei are considered to be
$\gamma$-soft \cite{Wu96,Bak78,Cas85} or transitional between
$\gamma$-soft and axially deformed nuclei, which is indicated by the
large values of the quadrupole moments of the $2^+_1$ states. As
examples for vibrational nuclei Cd and Pd nuclei are shown, and Gd and
Dy nuclei for the axially deformed case. We used $c_{0i}^N$ values
from the appropriate dynamical symmetry.

In Table \ref{tab:tkerne} the relations (\ref{eq:cdef}) are tested
with satisfactory overall agreement. Additionally, the values of
$K_4^{\rm appr.}$ are given in Table \ref{tab:tkerne}.

Table \ref{tab:qkerne} shows the experimental
values of $Q_{2^+_1}^2$ and $B(E2;0^+_{QQ} \rightarrow 2^+_1)$
for the chosen nuclei, compared to the values obtained by the
relations. The values of $Q_{2^+_1}^2$, obtained from the relations
(\ref{eq:q21aeb}) and (\ref{eq:q21beb}), agree with the experimental
values within the errors
in most cases. A high accuracy of data is necessary for significant
results, especially for the vibrator-like and the $\gamma$-soft
nuclei, for which the quadrupole moments become very small. 

As an example for discrepancies, we consider $^{188}$Os for which
the $B(E2;0^+_2\rightarrow 2^+_1)$ value is very small, as expected
for an $O(6)$ nucleus. The two-$Q$-phonon content of this state should 
therefore be very small. However, the values from
Eqs. (\ref{eq:q21aeb}),(\ref{eq:q21beb}) in Table \ref{tab:qkerne} may 
refer to a higher lying $0^+$ state with larger two-$Q$-phonon
contribution, for which the lifetime is not known. Thus, with the
missing $E2$ strength in $^{188}$Os, the value of $Q_{2^+_1}$ is
underestimated by relation (\ref{eq:q21beb}), while
Eq. (\ref{eq:q21aeb}) describes the quadrupole moment well.

One finds significant deviations for other nuclei, too. For example,
in $^{194}$Pt the large $B(E2;0^+_4\rightarrow 2^+_1)$ value indicates a
two-$Q$-phonon structure for the $0^+_4$ state, in contradiction with
the $O(6)$ prediction. Thus, this transition has been included in the
calculation of the $B(E2;0^+_{QQ}\rightarrow 2^+_1)$ value. The
value of $K_4^{\rm appr.}=0.8$ in 
$^{192}$Os is considerably smaller than its minimally allowed value:
1. This may be due to the small experimental value of
$B(E2;4^+_1\rightarrow 2^+_1)$. Also $K_4^{\rm appr.}$ for $^{108}$Pd
is unexpectedly small, which does not support the vibrational
character of this nucleus. In the Cd isotopes considered the measured
$Q_{2^+_1}$ are smaller than expected from the relations.

To summarize, we propose a simple method to derive sets of relations
between the experimentally observable reduced matrix elements of the
quadrupole operator. This approach is based on the use of the
quadrupole shape invariants, the selection rules of the $Q$--phonon
scheme and the fact that corrections from noncommutativity of the
components of the quadrupole moment operator in the IBM-1 are
small. As an example of the general
scheme, fourth order $Q$-invariants of the ground state are given. One
can apply the scheme also to higher order invariants, e.g. $q_5$ or
$q_6$, or to invariants built on excited states. The accuracy of the
derived relations is checked for finite boson number $N$ over the
whole parameter space of the ECQF-IBM-1 Hamiltonian and is shown to be
rather good. A satisfactory agreement between data and
theoretical relations has been obtained in many cases, but some
exceptions clearly need further study.

For fruitful discussions we thank A. Dewald,
A. Gelberg, J. Jolie, T. Otsuka and Yu.V. Palchikov.
One of the authors (P.B.) wants to thank the Institute for Nuclear
Theory at the University of Washington for its hospitality during the
final stages of this work, and R.J. thanks the Cologne University for
support. We thank K. Jessen for the careful reading of this paper.
This work was supported by the Deutsche
Forschungsgemeinschaft under Contract No. Br 799/10-1.

%%%%%%%%%%%%%%%%%%%%%%%% Tables %%%%%%%%%%%%%%%%%%%%%%%%

\begin{table}[htb]
\caption{Values of $t_4^{(0)}$, $1/c_02^N t_4^{(2)}$ and $1/c_04^N
t_4^{(4)}$ for various nuclei. The three values should agree according 
to Eq. \ref{eq:cdef}}
\label{tab:tkerne}
%\vspace{0.5cm}
\begin{center}
\begin{tabular}{cc|ccc|c}
 & data taken & $t_4^{(0)}$ & $1/c_{02}^N \cdot t_4^{(2)}$ & $1/c_{04}^N \cdot
t_4^{(4)}$ & $K_4^{\rm appr.}$ \\
 & from & $[$e$^2$b$^2]$ & $[$e$^2$b$^2]$ & $[$e$^2$b$^2]$ & \\
\hline
$^{186}$Os & \cite{Wu96} & $2.84(7)$ & $2.79(56)$ &
$3.06(17)$ & $1.06(3)$ \\[1mm]
$^{188}$Os & \cite{Wu96} & $2.52(3)$ & $2.72(48)$ &
$2.82(8)$ & $1.08(1)$ \\[1mm]
$^{190}$Os & \cite{Wu96} & $2.36(6)$ & $1.97(40)$ &
$2.28(16)$ & $0.93(3)$ \\[1mm]
$^{192}$Os & \cite{Wu96} & $2.12(3)$ & $2.20(30)$ &
$1.84(6)$ & $0.82(1)$ \\[1mm]
$^{194}$Pt & \cite{Wu96} & $1.56(12)$ & $1.96(12)$ &
$1.54(5)$ & $1.00(4)$ \\[1mm]
$^{196}$Pt & \cite{Bij80} & $1.34(6)$ & $1.53(25)$ & $1.56(9)$ &
$1.08(7)$ \\
\hline
$^{106}$Pd & \cite{Sve95} & $0.76(7)$ & $0.86(10)$ &
$0.83(9)$ & $1.19(9)$ \\
$^{108}$Pd & \cite{Sve95} & $0.92(11)$ & $1.10(13)$ &
$0.86(9)$ & $1.04(9)$ \\
$^{112}$Cd & \cite{Ra89,Ju80} & $0.65(5)$ & $0.37(6)$ &
$0.76(7)$ & $1.41(14)$ \\ 
$^{114}$Cd & \cite{Fa88} & $0.60(3)$ & $0.53(8)$ &
$0.77(5)$ & $1.39(12)$ \\ 
\hline
$^{156}$Gd & \cite{La83,Go81} & $4.67(13)$ & $4.58(23)$
& $4.66(13)$ & $0.98(3)$ \\ 
$^{158}$Gd & \cite{La83,Al88,Go81} & $5.03(15)$ &
$5.01(25)$ & $5.20(14)$ & $1.02(4)$ \\ 
$^{160}$Gd & \cite{Klu01,Ro77} & $5.25(5)$ & $5.36(26)$
& $5.20(13)$ & $0.98(2)$ \\ 
$^{164}$Dy & \cite{Go81} & $5.57(8)$ & $5.16(100)$ &
$5.12(27)$ & $0.91(5)$ \\ 
\end{tabular}
\end{center}
\end{table}

%\clearpage

\begin{table}
\caption{Comparison of the quadrupole moments of the $2^+_1$ state and 
the reduced transition strengths of the $0^+_{QQ}\rightarrow 2^+_1$
transition for various nuclei.}
\label{tab:qkerne}
%\vspace{0.5cm}
\begin{center}
\begin{tabular}{c|ccc|ccc}
 & \multicolumn{3}{c|}{$Q_{2^+_1}^2$} & 
\multicolumn{3}{c}{$B(E2;0^+_{QQ}\rightarrow 2^+_1)$} \\
 & $[$e$^2$b$^2]$ & $[$e$^2$b$^2]$ & $[$e$^2$b$^2]$ & $[$e$^2$b$^2]$ &
$[$e$^2$b$^2]$ & $[$e$^2$b$^2]$ \\[2mm]
 & Eq. (\ref{eq:q21aeb}) & exp. & Eq. (\ref{eq:q21beb}) &
Eqs. (\ref{eq:q21aeb},\ref{eq:q21beb}) & exp. & Eq. (\ref{eq:q21beb}) \\
\hline
$^{186}$Os & 1.97$^{+14}_{-29}$ & 1.76$^{+26}_{-44}$ &
1.80$^{+10}_{-12}$ & 0.25$^{+16}_{-17}$ & 0.040$^{+24}_{-16}$ &
$<$ 0.35 \\
$^{188}$Os & 1.80$^{+7}_{-21}$ & 1.72$^{+10}_{-38}$ &
1.56$^{+8}_{-8}$ & 0.30$^{+8}_{-8}$ & 0.0061$^{+3}_{-3}$ &
0.20$^{+15}_{-20}$ \\
$^{190}$Os & 1.14$^{+15}_{-30}$ & 0.90$^{+19}_{-32}$ &
1.20$^{+9}_{-8}$ & $<$ 0.10 & 0.014$^{+2}_{-2}$ &
0 \\
$^{192}$Os & 0.56$^{+6}_{-19}$ & 0.84$^{+24}_{-8}$ &
0.78$^{+7}_{-8}$ & 0 & 0.004$^{+1}_{-1}$ &
0.08$^{+32}_{-8}$ \\
$^{194}$Pt & 0 & 0.20$^{+2}_{-7}$ &
$<$ 0.01 & 0.08$^{+6}_{-8}$ & 0.100$^{+6}_{-6}$ &
0.50$^{+9}_{-17}$ \\
$^{196}$Pt & $0.26(9)$ & $0.24(18)$ & $0.10(8)$ & $0.24(11)$ &
$0.02(1)$ & $0.21(26)$ \\
\hline
$^{106}$Pd & $0.28^{+8}_{-22}$ & $0.30^{+5}_{-6}$ & $0.23^{+7}_{-7}$
& $0.20^{+11}_{-11}$ & $0.14^{+2}_{-2}$ & $0.23^{+11}_{-11}$  \\
$^{108}$Pd & $0.20^{+9}_{-20}$ & $0.38^{+4}_{-8}$ & $0.24^{+10}_{-8}$
& $0.11^{+11}_{-11}$ & $0.16^{+2}_{-2}$ & $0.35^{+11}_{-17}$  \\
$^{112}$Cd & $0.43(6)$ & $0.14(3)$ & $0.35(5)$ & $0.27(8)$ & $0.16(5)$
& 0  \\
$^{114}$Cd & $0.31(4)$ & $0.13(6)$ & $0.18(3)$ & $0.26(6)$ & $0.090(5)$
& $0.02(9)$  \\
\hline
$^{156}$Gd & $3.79(11)$ & $3.72(15)$ & $3.79(15)$ & $< 0.18$ & n.a.
& $< 0.18$  \\
$^{158}$Gd & $4.19(11)$ & $4.04(16)$ & $4.05(18)$ & $0.17(20)$ & n.a.
& $< 0.28$  \\
$^{160}$Gd & $4.20(10)$ & $4.33(17)$ & $4.23(14)$ & $< 0.09$ & n.a.
& $0.11(26)$ \\
$^{164}$Dy & $4.09(22)$ & $4.12(81)$ & $4.46(15)$ & 0 & n.a. 
& $< 0.60$ \\
\end{tabular}
\end{center}
\end{table}

%%%%%%%%%%%%%%%%%%%%%%%% Figures %%%%%%%%%%%%%%%%%%%%%%%%
%\onecolumn

\end{document}